\documentclass[aps,pre,preprint]{revtex4-1}          

\usepackage{graphicx} 
\usepackage{subfigure} 

\begin{document}

\title{One-dimensional linear array of cylindrical posts for size-based deterministic separation of binary suspensions}

\author{Raghavendra Devendra\textit{$^{a, b}$ and German Drazer\textit{$^{a, b}$}}}
\affiliation{\textit{$^{a}$~Department of Chemical and Biomolecular Engineering, Johns Hopkins University, Baltimore MD 21218 USA.}}
\affiliation{\textit{$^{b}$~Present Address: Department of Mechanical and Aerospace Engineering, Rutgers, The State University of New Jersey, Piscataway NJ 08854 USA. E-mail: german.drazer@rutgers.edu}}

\begin{abstract}
We investigate the motion of suspended particles past a single line of equally spaced cylindrical posts that is slanted with respect to the driving force.
We show that such a one-dimensional array of posts can fractionate particles according to their size, with small particles permeating through the line of posts but larger particles being deflected by the steric barrier created by the posts, even though the gaps between posts are larger than the particles.
We perform characterization experiments driving monodisperse suspensions of particles of different size past the line of posts over the entire range of 
forcing orientations and present both the permeation probability through the individual gaps between the posts as well as the fraction of permeating
particles through the one-dimensional array.  In both cases, we observe a sharp transition from deflection to permeation mode that is a function of particle size, thus enabling separation.  We then drive binary mixtures at selected orientations of the line of posts and demonstrate high purity and efficiency of separation.
 \end{abstract}

\maketitle

\section{Introduction}

An important unit operation in a variety of lab on a chip systems is the separation of suspended mixtures of species. 
As a result, a number of microfluidic devices have been created either to miniaturize conventional separation techniques or to implement novel separation strategies. \cite{pamme2007,Li2007,Lenshof2010,Stone2004}  Different aspects of the separation process are emphasized depending on the application, but a common trend is the development of continuous separation processes, which can be readily integrated with downstream operations in a lab on a chip platform.\cite{Bernate2011,2012-Jorge-PRL,Sai2006,Devendra2012,Pamme2004,Kulrattanarak2008,yamada2005,Yamada2006,hsu2008,Halli2009,Huh2007}

Deterministic lateral displacement (DLD) is a particularly promising, two-dimensional continuous separation method to fractionate a mixture of suspended species.\cite{Huang2004,Heller2008}  In this technique, components of different size migrate in different directions as they are driven through a periodic array of cylindrical posts.  DLD has been successfully implemented for the fractionation of diverse samples.\cite{Inglis2006,Holm2011,Green2009,Loutherback2010,Long2008,Davis2006}  In previous work, we investigated the mechanisms leading to separation in DLD and demonstrated an alternative operation mode, in which suspended species are driven through a two-dimensional (2D) array of posts by a constant force, specifically gravity (g-DLD). \cite{Frechette2009,Balvin2009,herrmann2009,koplik2010,Devendra2012,2011-Ming-APL,Bowman2012,2013-Sumedh-JFM}
The motion of suspended particles in DLD shows directional locking, in which the migration angle remains constant and equal to a lattice direction over a finite range of forcing orientations.\cite{Frechette2009} In particular, for small enough angles between the external force and a line of posts corresponding to a column (or a row) in a square array, the particles move down a lane between two adjacent columns (or rows) without crossing them.  Previous experiments also established that, as the forcing angle increases (from zero), there is a critical transition angle, above which the particles are able to move across columns in the array. \cite{Devendra2012} This critical transition angle depends on particle size, thus allowing for separation. Moreover, the highest resolution of separation in a binary mixture was found to occur near the critical transition angle.\cite{Devendra2012}  These results suggest that a single line or a one-dimensional (1D) array of cylindrical posts in lieu of the entire 2D array, could in principle, be used to fractionate a mixture of particles. 

Here, motivated by the results discussed above, we investigate the motion of suspended particles driven past a slanted line of uniformly spaced cylindrical posts by a constant external force.  We demonstrate that such a 1D array of posts can, in fact, fractionate a mixture of suspended particles with high purity and efficiency.  More specifically, we shall show that, depending on the angle between the line of posts and the external force, large particles will be deflected sideways whereas small particles will permeate through the 1D array of posts.  It is important to note that the gap between the posts is larger than the largest particles used here. On the other hand, the deflection of particles of a given size is only possible when the {\it projected gap}, that is the gap between posts projected in the direction of the driving force, is smaller than the particles.  This introduces a different approach to separation based on a steric barrier (the 1D array of posts) that is different from a traditional membrane or filter.

\section{Materials and methods}

\subsection{Device fabrication and experimental setup}

\begin{figure}[!htb]
\centering
  \includegraphics[trim=3cm 6cm 3cm 9cm, clip=true, width=10 cm]{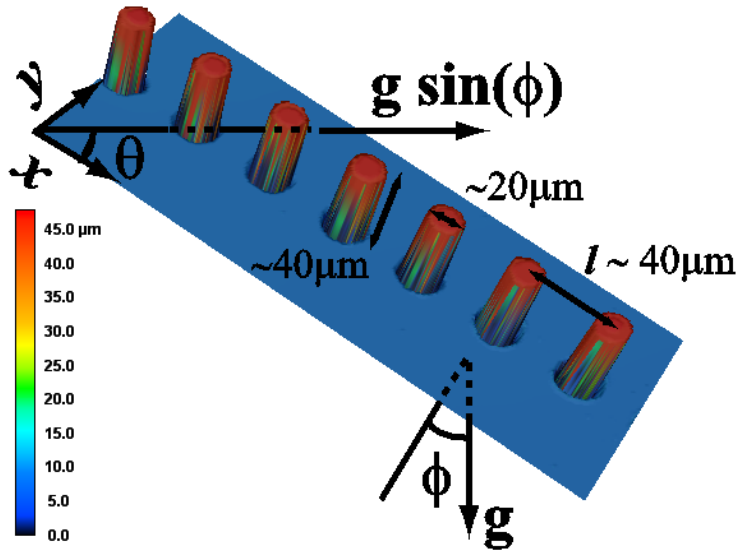} 
  \caption{Three-dimensional image of the device indicating the microscope tilt angle $\phi$, the forcing angle $\theta$, diameter of the posts $2R \approx 20 \mu$m, the spacing between the posts $\ell=40\mu$m and the height of the posts $\approx40\mu$m.}
  \label{fgr:ColumnRescaled}
\end{figure}

The devices were fabricated using photolithography in a clean room.  A negative photoresist (SU8-3025) was spun coated on top of a standard microscope glass slide, exposed to UV light and developed to obtain a line of cylindrical posts on the glass slide.  The thickness of the photoresist and therefore, the height of the resulting posts, was approximately 40${\mu}$m.  The profile of a fabricated line of posts obtained with a 3D Laser Scanning Microscope VK-X100/X200 (Keyence Corp., Japan) is shown in  Figure \ref{fgr:ColumnRescaled}.  The measured diameter of the posts is $2R = 19.5 \mu$m and the centre-to-centre separation is $\ell =40\mu$m.  Several of these lines of posts were fabricated on a single glass slide.  Individual lines of posts were separated by 1000$\mu$m from each other to avoid any effect on the particles moving past a given line from a neighbouring line.  The lines were used singularly in independent experiments.  The diameter of the posts and the spacing between them were designed to be comparable to the diameter of the largest particles used in the experiments.   

The lines of cylindrical posts were surrounded by double sided adhesive tape (Grace Bio-Labs, Inc., OR) to create a containing well on the glass slide.  We performed experiments using silica particles with a density of 2 g/cm$^3$ and average diameter of 4.32 ${\mu}$m (Bangs Laborarties, Inc., CA), 10${\mu}$m, 15${\mu}$m and 20${\mu}$m (Corpuscular Inc., NY).  The particles were suspended in 0.1 mM aqueous KOH solution (to prevent sticking).  They were then introduced into the well and allowed to settle down at the bottom of the device.  Without trapping any air inside, the well was sealed using a glass cover-slip (and double sided adhesive tape) to eliminate the possibility of convective effects.  The device containing the suspended particles was then mounted on a microscope and the microscope was tilted at an angle, $\phi=16^\circ$, thus exploiting gravity to drive the particles down the slide and past the line of cylindrical posts.  More details on the experimental procedure can be found in reference \citenum{Devendra2012}. 
     
\subsection{Image acquisition and data analysis}

Images were captured at a frame interval of 2s in bright field, using a Rolera Mgi+ EMCCD camera (Q-imaging, BC) and IP lab software (Biovision Tech, PA).  We used manual tracking plugin (by Fabrice Cordelieres) in ImageJ (NIH, MD) to track the particles.  The field of view consisted of 21 posts.  The stream of particles was not focussed from a single starting point and as a result, the first interacting post for a given particle could be any one of the posts in the field of view.  The forcing angle,  ${\theta}$, is defined as the angle between the direction of the driving force (gravity projected onto the plane of the device i.e. $xy$-plane in Figure \ref{fgr:ColumnRescaled}) and the line of posts ($x$-axis in Figure \ref{fgr:ColumnRescaled}) and was measured by tracking particles in an (open) area free of posts.  Experiments were performed independently at forcing angles with no particular increasing or decreasing order to avoid any cumulative error or correlation between the experiments.  We only considered particles that travelled at least 100${\mu}$m in the driving direction to reduce any fluctuations in the value of the measured forcing angle.  The particles moving in the open area and those interacting with the posts were tracked separately.  Approximately, 20 particles were tracked in each experiment for a given forcing angle and a given particle size.
In order to minimize the effect of polydispersity in the supplied particles, especially 10${\mu}$m, 15${\mu}$m and 20${\mu}$m particles, we used calibration scales during tracking and considered only those particles within 10\% of the specified size.  In order to capture the behaviour in the dilute limit, the trajectories of particles that interact with another particle were discarded.  On occasions, where a large particle blocked a spacing between the posts, the trajectories that interacted with such stuck particles were split into separate trajectories by cropping out the periods during which the interactions with the stuck particles took place.  In rare instances, if a post was significantly different in size and shape from the design, the particles that interacted with that specific post were eliminated from the analysis.

\section{Results and discussion}

The proposed separation approach is based on the hypothesis that we can use a single 1D array of cylindrical posts to fractionate a suspension of particles of different size into two streams.  The main assumption is that, for a given particle size (and material), there is a critical forcing angle $\theta_c$, that characterizes their motion past the line of posts.  Specifically, particles would be displaced laterally by the line of posts for $\theta<\theta_c$ (\textit {deflection mode}), but would pass through the line of posts for $\theta>\theta_c$ ({\it permeation mode}). The second assumption is that, as observed in previous DLD experiments using 2D arrays of posts, the critical angle depends on particle size, thus leading to size-based separation.\cite{Devendra2012} In particular, a polydisperse suspension of particles moving past a line of posts oriented at an angle $\theta$, would result in two distinct streams of particles: one permeating through, composed of particles for which $\theta_c < \theta$ and the other deflecting along the line of posts, composed of all the particles for which $\theta_c > \theta$. Moreover, previous experiments in 2D arrays suggest that the critical forcing angle is an increasing function of particle size and therefore, there would be a critical size $a=a_c$, where $a$ is the particle radius, such that $\theta_c(a_c)=\theta$ and smaller particles ($a<a_c$) would simply permeate through the 1D array of posts but larger particles ($a>a_c$) would be deflected.  

\subsection{Probability analysis}

In order to investigate the postulated critical behaviour, we analyze the motion and interaction of the particles with individual posts as independent stochastic events in a Bernoulli process.  An event here is specific to a particle in a given gap between two consecutive posts.  It is defined as a {\it success} if the particle passes through the gap and a {\it failure} 
if the particle does not pass through the gap but gets laterally displaced and continues towards the next gap downstream. Then, we define the probability of crossing, $p$, as the frequency ratio $\nu_c/n$, where $\nu_c$ is the number of successes and $n$ is the total number of events. We also estimate the uncertainty in the determination of the probability of crossing with the variance $\sigma=\sqrt{\frac{p(1-p)}{n}}$. We characterize the downstream output by defining a permeation fraction, $r_p=\nu_c/N$, where $N$ is the total number of particles interacting with the line of posts. The uncertainty in the determination of the permeation fraction is estimated with the variance $\sigma=\sqrt{\frac{p(1-p)}{N}}$.  In terms of the experimental variables, $\theta_c$ is estimated as the average of lower bound $\left( \theta_c^L \right)$ and upper bound $\left( \theta_c^U \right)$ values of the critical transition angle, defined by the highest measured forcing angle corresponding to $p=0$ and the lowest measured forcing angle corresponding to $p=1$, respectively.  

\subsection{Experimental results}

\begin{figure}[ht]
\centering
  \includegraphics[trim=0cm 9cm 0cm 9cm, clip=true, width=0.6\textwidth]{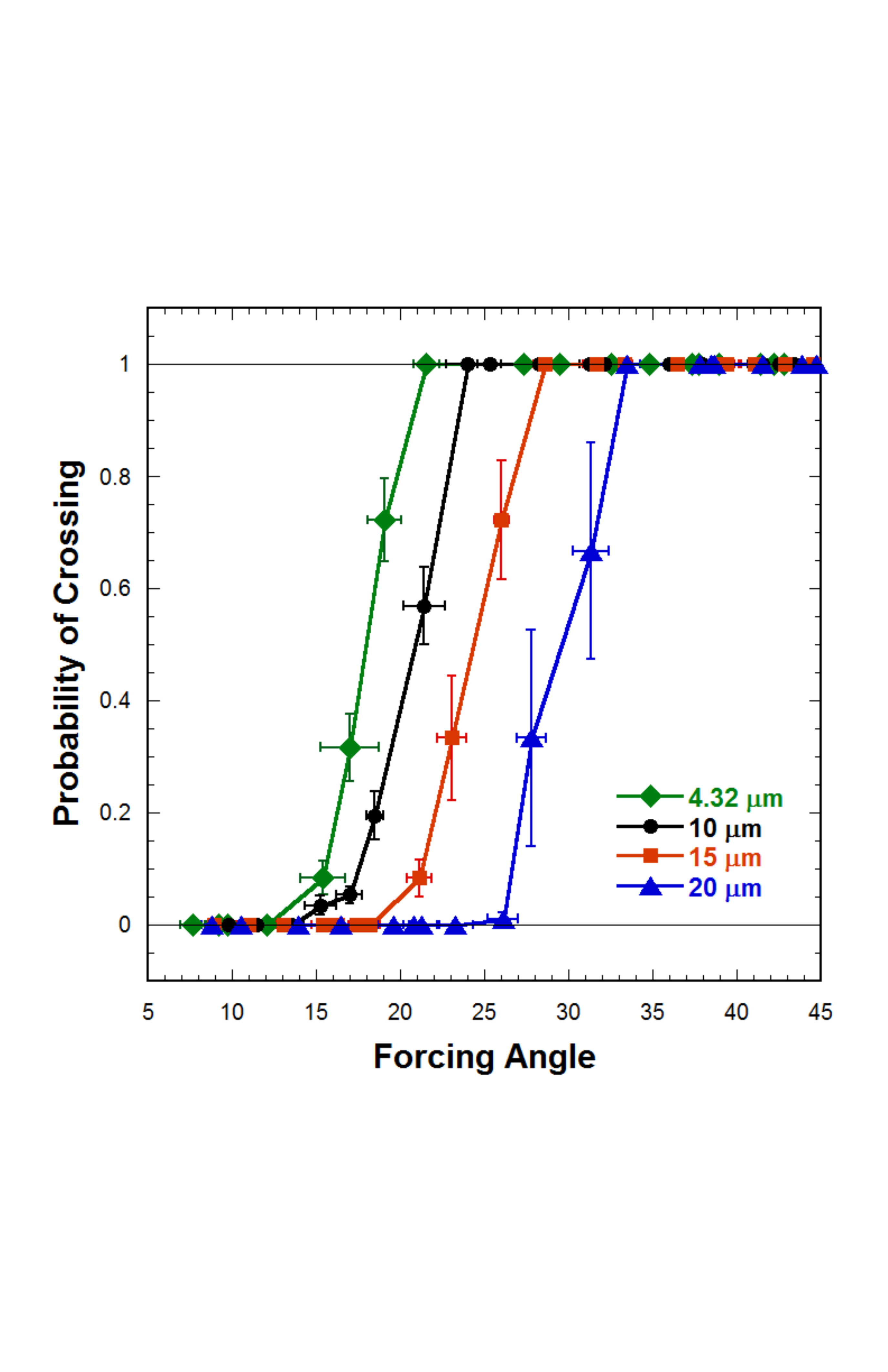}
  \caption{Probability of crossing, $p$, for $4.32\mu$m, $10\mu$m, $15\mu$m and $20\mu$m particles as a function of forcing angle.}
  \label{fgr:IndependentobstacleSi}
  \end{figure}

In Figure \ref{fgr:IndependentobstacleSi}, we present the probability of crossing, $p$, as a function of the forcing angle, ${\theta}$, for all particle sizes and the entire range of forcing directions.  Each probability value is obtained from an independent experiment at the corresponding forcing angle.  In all cases, we observe that the value of the probability is initially zero for a range of forcing directions and sharply transitions to $p=1$ over a small range of forcing angles.  This behaviour is consistent with the directional locking observed in the 2D arrays\cite{Devendra2012} and with the existence of a critical angle at which the motion of a given size of particles transitions from deflection to permeation, as discussed above.  It is also clear in Figure \ref{fgr:IndependentobstacleSi} that the critical angle is different for particles of different size, thus indicating the potential for using a single 1D array of posts for separation.  Both $\theta_c^L$ and $\theta_c^U$ increase with the size of the particles.  The measured values are shown in Table \ref{tbl:angledata}, where we also estimate $\theta_c$ as the average between $\theta_c^L$ and $\theta_c^U$.  The estimated critical angle also shows the same trend as in 2D arrays.\cite{Devendra2012}

\begin{table*}
\small
\caption{\ Critical forcing angles for all particles in single species experiments}
\label{tbl:angledata}
\begin{tabular*}{\textwidth}{@{\extracolsep{\fill}}lllllll}

Particle diameter & $\theta_c^L$ & $\theta_c^U$ & $\theta_c$ \\
\hline
4.32$\mu$m & $14^\circ\pm3^\circ$ & $20^\circ\pm2^\circ$ & $17^\circ\pm4^\circ$ \\
10$\mu$m& $14^\circ\pm2^\circ$ & $23^\circ\pm3^\circ$ & $19^\circ\pm4^\circ$\\
15$\mu$m& $20^\circ\pm1^\circ$ & $27^\circ\pm1^\circ$ & $23^\circ\pm2^\circ$ \\
20$\mu$m & $25^\circ\pm2^\circ$ & $32^\circ\pm2^\circ$ & $29^\circ\pm4^\circ$ \\
\hline
\end{tabular*}
\end{table*}

\begin{figure}[ht]
\centering
  \includegraphics[trim=0cm 9cm 0cm 9cm, clip=true, width=0.6\textwidth]{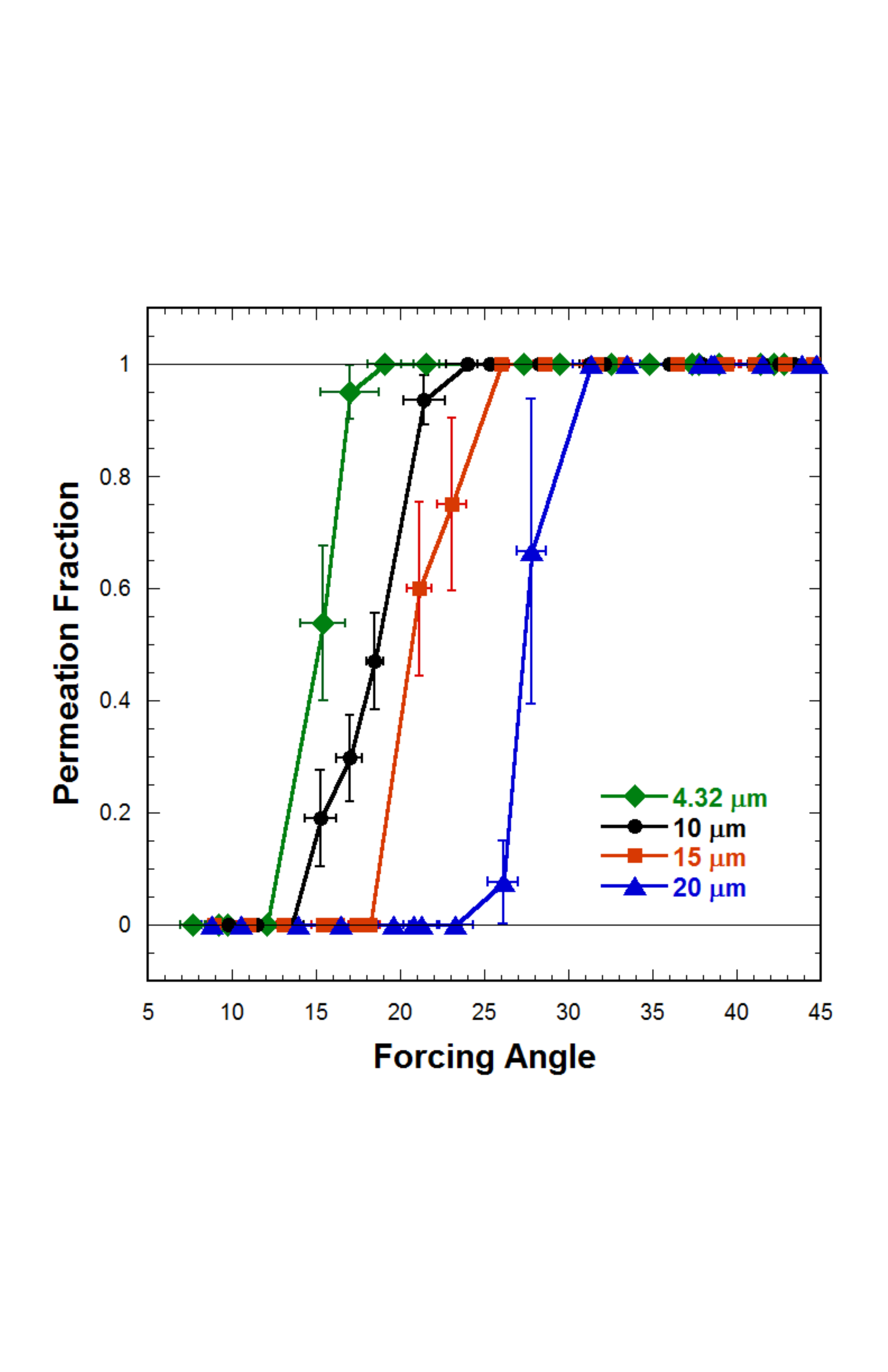}
  \caption{Permeation fraction, $r_p$, for $4.32\mu$m, $10\mu$m, $15\mu$m and $20\mu$m particles as a function of forcing angle.}
  \label{fgr:IndependentColumnSi}
  \end{figure}

The actual separation however, is better characterized by the  fraction of particles that permeates through the line of posts, that is the permeation fraction, $r_p$.  
In Figure \ref{fgr:IndependentColumnSi}, we present ${\it r_p}$ as a function of the forcing angle, ${\theta}$, for all particle sizes.  The transition from deflection to permeation is steeper, by definition of $r_p$, with the permeation angles at which $r_p=1$ typically reduced by $\approx 2^\circ$ compared to $\theta_c^U$.  
On the other hand, there is little change in the critical transition angles or the first angles for which the permeation fraction assumes a finite value, compared to $\theta_c^L$. 

\subsection{Binary experiments and Separability}

\begin{figure}[!htb]
\begin{center}

	\subfigure[]
	{%
		\label{fig:4and15sep}
		\includegraphics[trim=2cm 5cm 2cm 5cm, clip=true, width=0.45\textwidth]{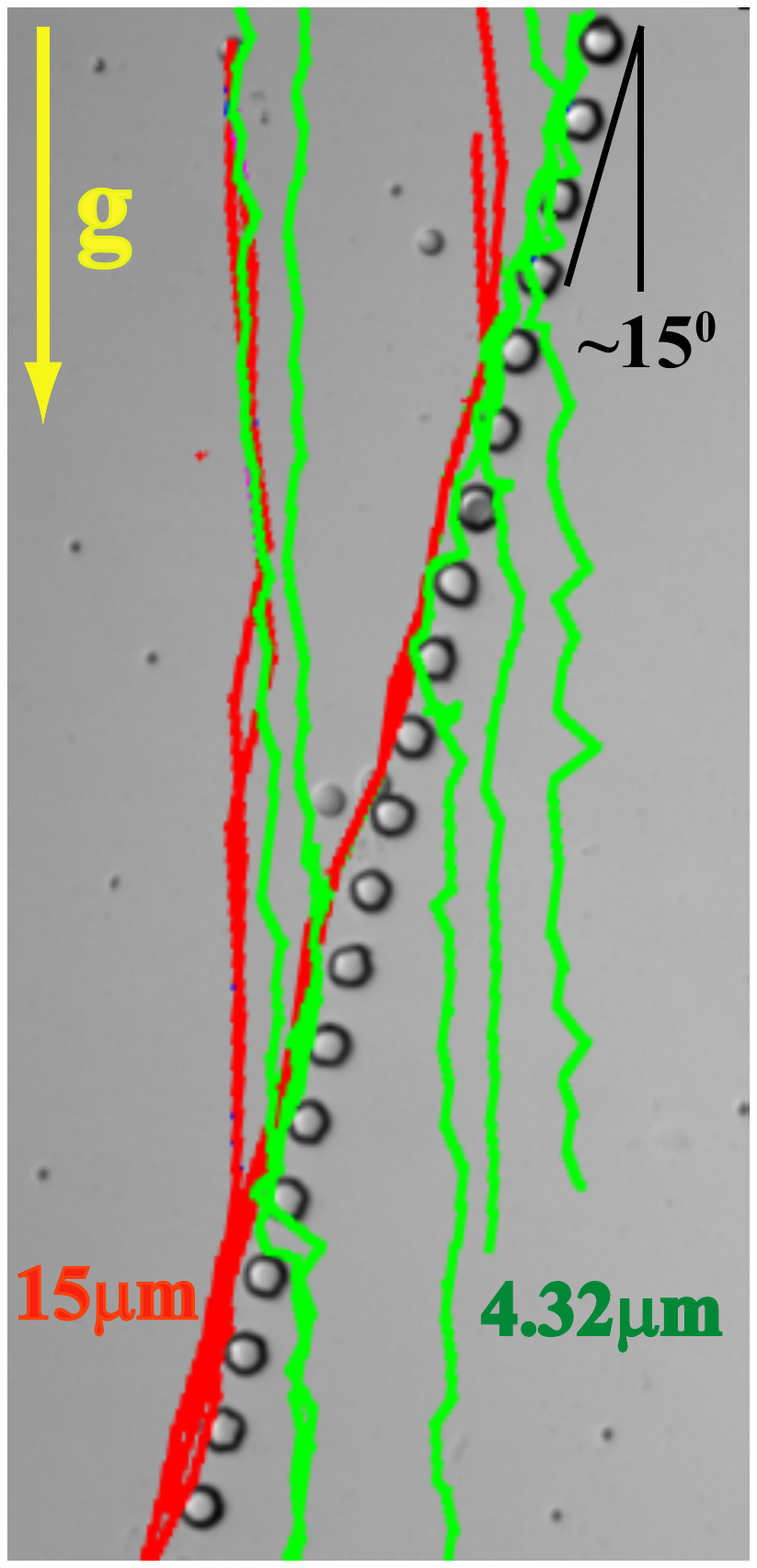}
	}%
	\subfigure[]
	{%
           	\label{fig:10and20sep}
           	\includegraphics[trim=2cm 5cm 2cm 5cm, clip=true, width=0.45\textwidth]{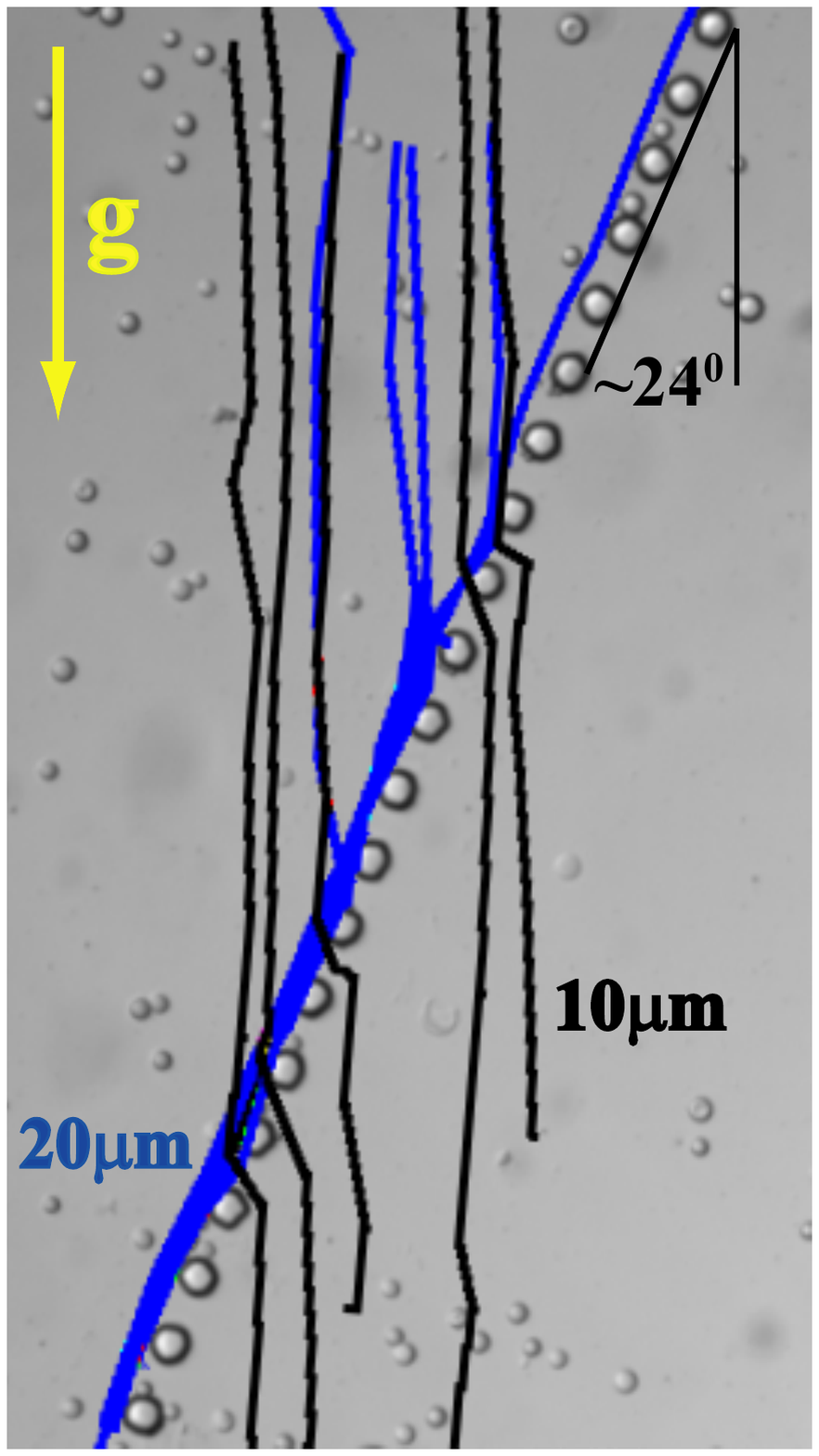}
        	}%
\end{center}
\caption
	{%
        	Representative trajectories during the fractionation of two different binary mixtures.  In both cases, we see permeation of the smaller particles in the mixture and deflection of the larger ones. (a) A mixture of $4.32 \mu$m and $15 \mu$m particles.  $\theta=15^\circ$.  (b) A mixture of $10 \mu$m and $20 \mu$m particles.  $\theta=24^\circ$.
     	}%
   \label{fig:Separation}

\end{figure}

Based on the results discussed above, obtained for different species separately, we selected specific forcing directions to investigate the separation of binary mixtures of $4.32\mu$m and $15\mu$m particles and of $10\mu$m and $20\mu$m particles.  The purpose is to identify specific forcing directions where smaller particles permeate through the line of posts while the larger particles deflect.  In Figure \ref{fig:Separation}, we present particle trajectories from two representative examples of separation in binary mixtures.  
In Figure \ref{fig:4and15sep}, the forcing direction is $\theta\approx15^\circ$ and the $4.32\mu$m particles exhibit a finite permeation fraction ($0<r_p<1$), whereas the $15\mu$m particles are completely deflected along the direction of the line of posts ($r_p=0$).  In Figure \ref{fig:10and20sep}, the forcing direction is $\theta\approx24^\circ$ and all the $10\mu$m particles permeate through the posts ($r_p=1$), while all the $20\mu$m particles are completely deflected ($r_p=0$).  These figures graphically illustrate the probability analysis discussed above, as well as the proposed mechanism for separation.

\begin{figure}[ht!]
\begin{center}
	\subfigure[]
	{%
		\label{fig:4and15BinaryObstacleSi}
		\includegraphics[trim=0cm 9cm 0cm 9cm, clip=true, width=0.4\textwidth]{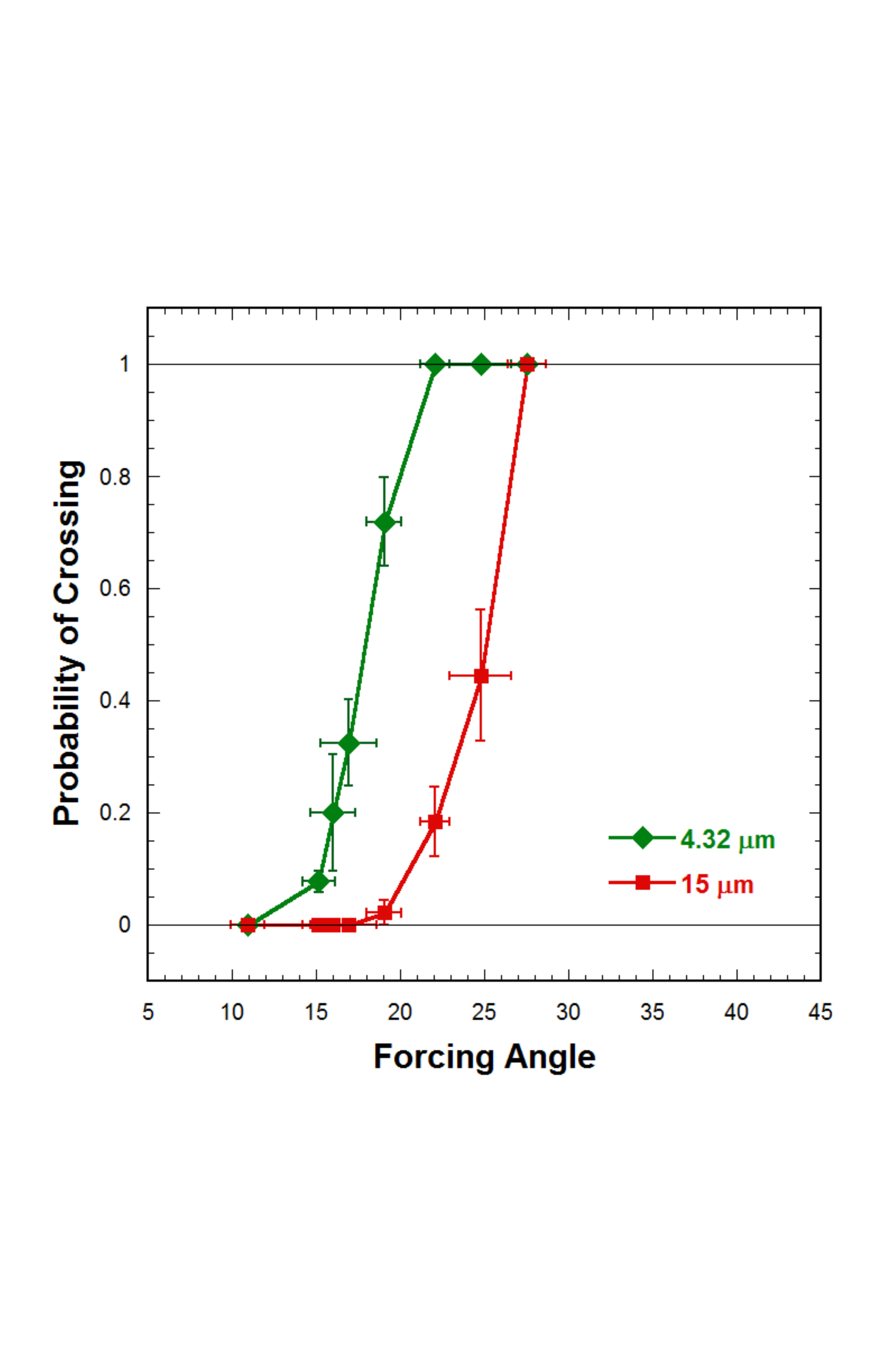}
	}%
	\subfigure[]
	{%
           	\label{fig:4and15BinaryColumnSi}
           	\includegraphics[trim=0cm 9cm 0cm 9cm, clip=true, width=0.4\textwidth]{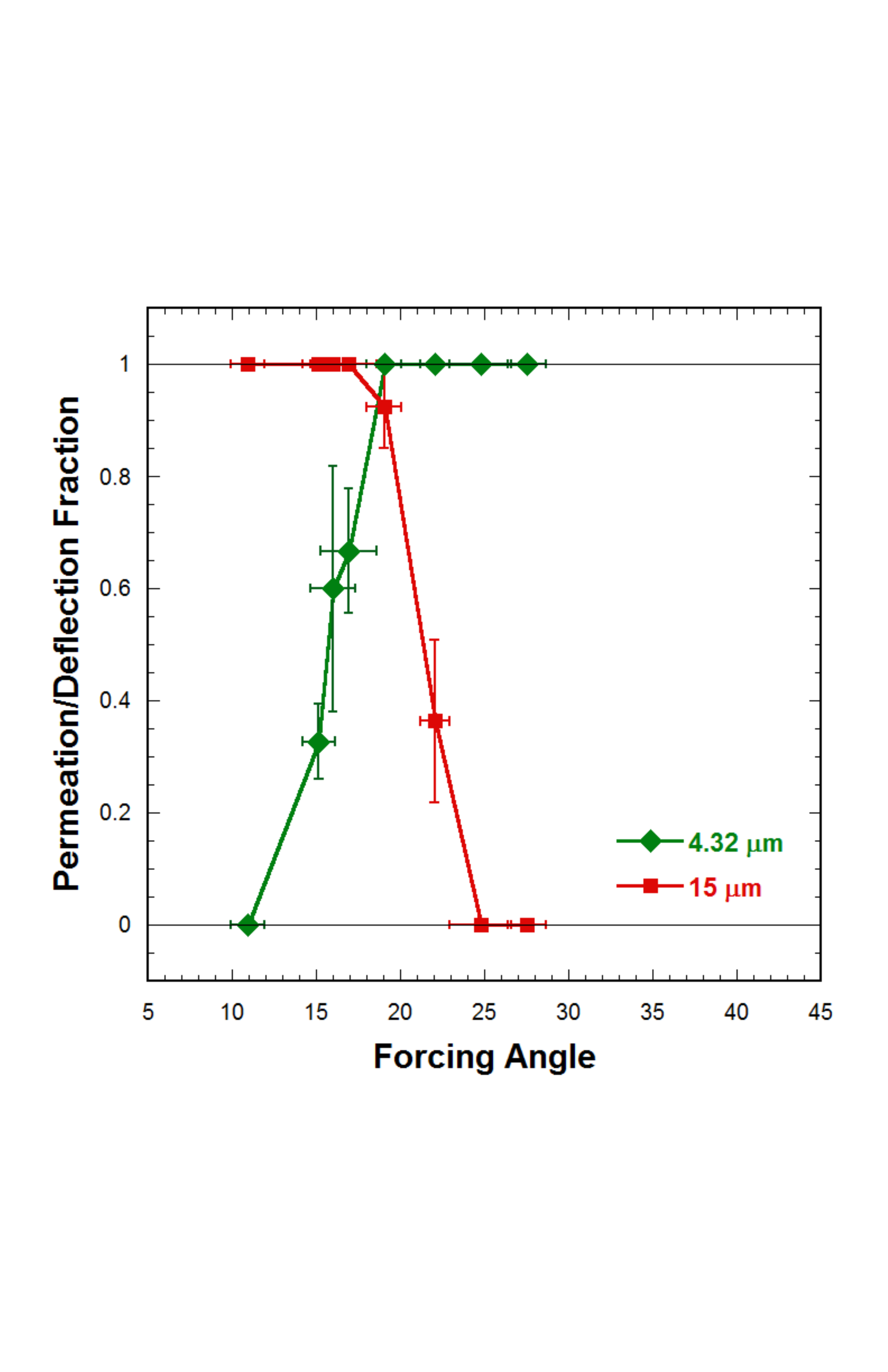}
        	}\\ 
       	\subfigure[]
	{%
            	\label{fig:10and20BinaryObstacleSi}
            	\includegraphics[trim=0cm 9cm 0cm 9cm, clip=true, width=0.4\textwidth]{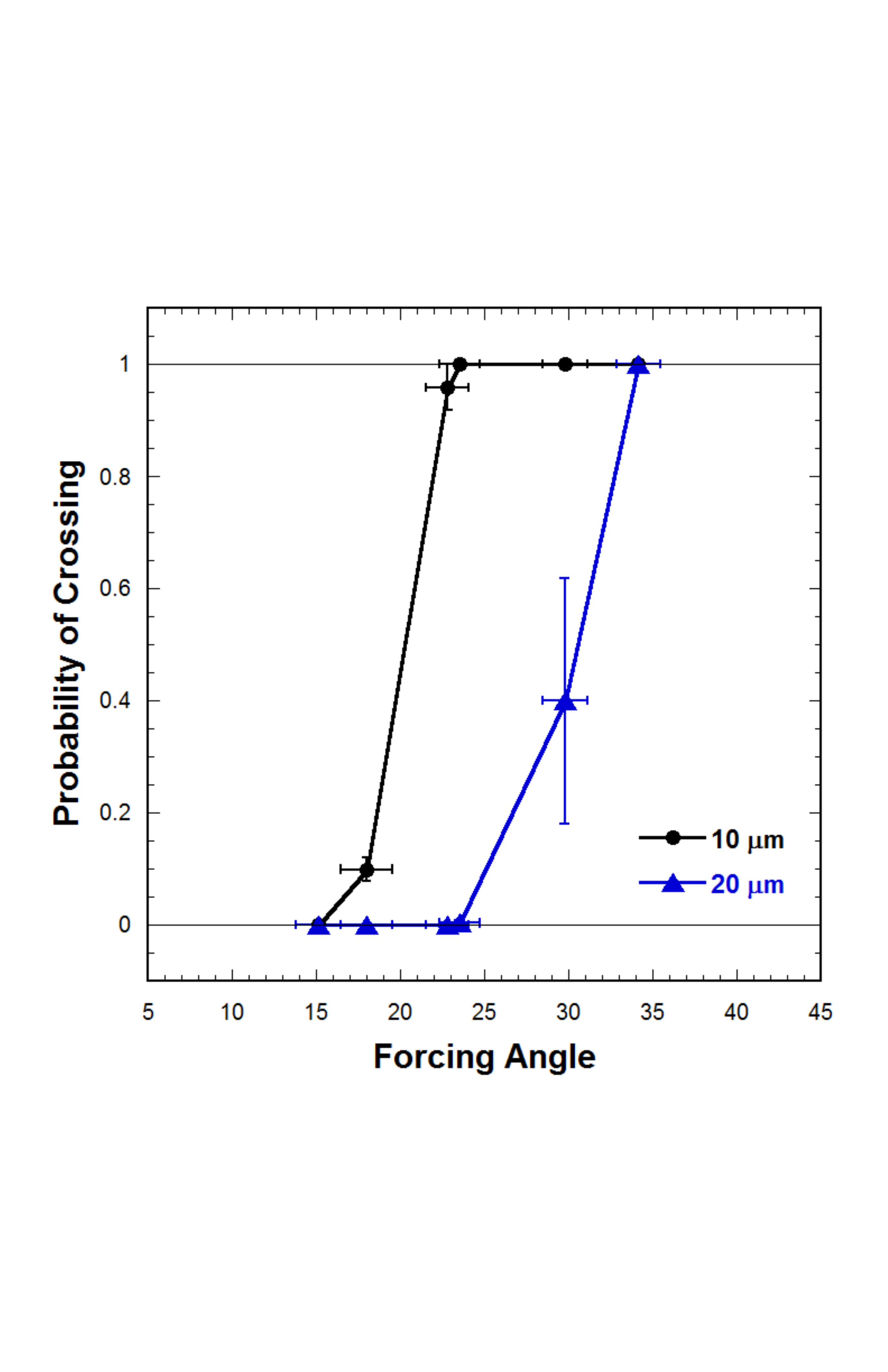}
        	}%
        	\subfigure[]
	{%
            	\label{fig:10and20BinaryColumnSi}
            	\includegraphics[trim=0cm 9cm 0cm 9cm, clip=true, width=0.4\textwidth]{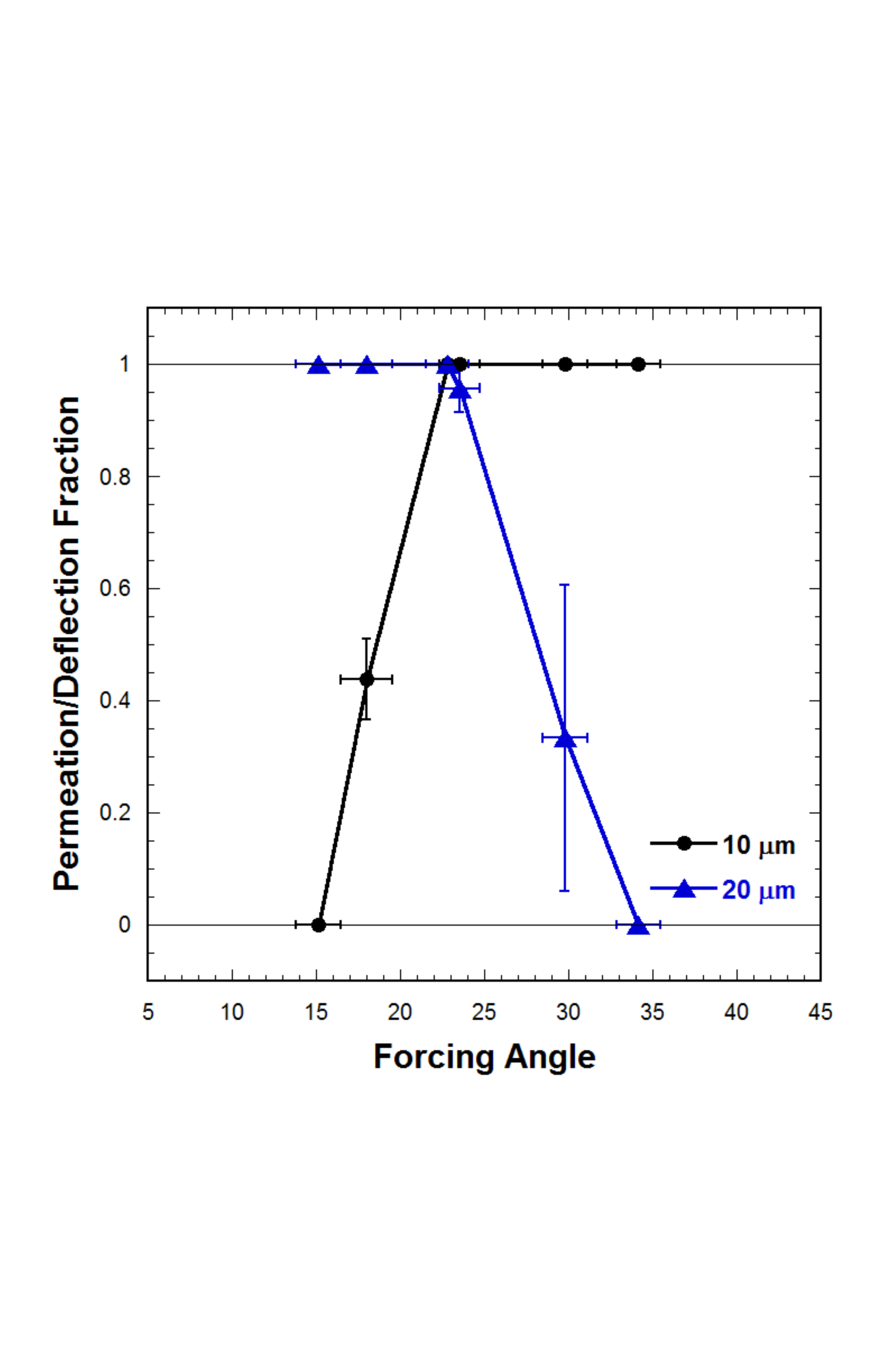}
        	}%
    \end{center}
    \caption
	{%
        	Probability of crossing for a binary mixture of (a) $4.32 \mu$m and $15 \mu$m and of (c) $10 \mu$m and $20 \mu$m particles.  
Permeation fraction for $4.32 \mu$m and $10 \mu$m particles and deflection fraction for $15 \mu$m and $20 \mu$m particles in a binary mixture of (b) $4.32 \mu$m and $15 \mu$m and of (d) $10 \mu$m and $20 \mu$m particles.
     	}%
   \label{fig:Binary}
\end{figure}

In Figure \ref{fig:Binary}, we present the probability of crossing at the selected forcing directions, both in a binary mixture of $4.32 \mu$m and 15 $\mu$m particles (Figure \ref{fig:4and15BinaryObstacleSi}) and of 10 $\mu$m and 20 $\mu$m particles (Figure \ref{fig:10and20BinaryObstacleSi} ). 
We also present the permeation fraction, $r_p$, of the smaller particle and the deflection fraction, $r_d=1-r_p$, of the larger particle in both binary mixtures in Figures \ref{fig:4and15BinaryColumnSi} and \ref{fig:10and20BinaryColumnSi}.  
High purity and efficiency of separation can be achieved in both binary mixtures at forcing direction for which, both $r_p$ and $r_d$ are close to $100\%$. 
For example, from Figure \ref{fig:4and15BinaryColumnSi}, we see that the optimum forcing angle to fractionate $4.32 \mu$m and 15 $\mu$m particles is around $\theta \approx 18^\circ$.  Similar estimation can be made from Figure \ref{fig:10and20BinaryColumnSi} for the separation of 10 $\mu$m and 20 $\mu$m particles, with the optimum forcing direction around $\theta \approx 23^\circ$.

\section{Conclusions}

We presented a simple and novel method for continuously separating components in a binary mixture using a single 1D array of cylindrical posts.  We initially performed characterization experiments, by driving monodisperse suspensions through a line of posts over the entire range of forcing orientations and analyzed the results in terms of the probability of a particle to permeate through the line of posts.  In each case, we observed a sharp increase in the permeation probability over a small range of forcing angles, indicating a transition from deflection to permeation mode as the forcing angle increases from zero.  We also showed that the critical transition angle increases with particle size. We finally used these characterization experiments to fractionate binary mixtures at selected force orientations,
showing excellent purity and efficiency of separation at angles close to the critical transition angle of the larger particles.

Let us finally note that, the probability of a particle to permeate through the line of posts depends on the number of posts.  It is expected that a longer line of posts increases the probability of crossing and thus, the permeation fraction.  Hence, increasing the number of posts results in a shift towards lower values of the forcing angle at which all particles permeate \emph{i.e.} smaller $\theta_c^U$.  On the other hand, small variations in the size and shape of posts, as well as other effects such as Brownian motion of the suspended particles, could induce early transitions and lower the purity (or efficiency) of the separation.  In this sense, a 1D array of posts is more sensitive to defects than a 2D DLD array.  Therefore, one should treat the number of posts as a variable used to optimize separation quality or resolution.    

The proposed separation approach has similarities with cross-flow fractionation, but it is important to note that in this case the size of the {\it pores} is larger than the size of the suspended particles and the system is thus less prone to clogging. Finally, let us mention that this method could be enhanced into multi-component separation by positioning consecutive lines of posts at increasing angles with respect to the force (or other arrangements).

\section*{Acknowledgements}
The authors acknowledge technical assistance in particle tracking by Roberto Passarro and Tsung-Chung Feng.  This material is based upon work partially supported by the National Science Foundation under grant CBET-0954840.

\end{document}